\documentclass[12pt]{article}%
\usepackage{eurosym}
\usepackage{amssymb}
\usepackage{amsfonts}
\usepackage{amsmath}
\usepackage{graphicx,epsf}
\usepackage{enumerate}
\usepackage{url}%
\usepackage{graphicx}
\providecommand{\U}[1]{\protect\rule{.1in}{.1in}}
\RequirePackage[colorlinks,citecolor=blue,urlcolor=blue]{hyperref}

\numberwithin{equation}{section}
\numberwithin{theo}{section}
\numberwithin{lemma}{section}
\numberwithin{coro}{section}
\numberwithin{example}{section}
\numberwithin{rem}{section}
\numberwithin{proposition}{section}
\begin{document}

\date{}
\title{Reclaiming the "frequentist" role of marginal likelihood in Bayesian belief revision}
\author{Abdelhakim~Aknouche\\Department of Applied Statistics, Stony Brooke Institute at \\Anhui University (SBIAU), Hefei, China}
\maketitle

\begin{abstract}
In modern Bayesian computation and parametric estimation, the marginal
likelihood, serving as the denominator $P(D)$ in Bayes' Theorem, is routinely
bypassed via unnormalized proportionality relations. Even within specialized
model-selection frameworks where it is explicitly evaluated to compute Bayes
Factors, the denominator is treated purely as a static constant. This note
evaluates a subtle analytical oversight resulting from this computational
convenience. Through the analysis of a simplified, sequential
partial-information system, we show that the marginal probability possesses a
critical dual layer of information: while the posterior probability determines
the local magnitude of a belief update upon a solitary trial, the marginal
denominator governs the physical, long-run frequentist cadence of that update
across a historical horizon. Discarding the denominator computes what an
observer ought to believe once a specific dataset manifests, but erases the
data-generating reality that governs how frequently that inferential state
occurs in nature. We propose reclaiming the marginal likelihood as an active,
real-time regularizer. We introduce three diagnostic measures to regulate
recursive online estimation gains, and construct valid mixture probabilities
that blend the prior and posterior to function as surprise-activated or
conservative regulators. This paired probability framework may offer a robust
regularizing mechanism for sequential estimation architectures and
quantitative risk management scenarios under non-stationary distribution shifts.


\end{abstract}

\section{Introduction: The computational dismissal of $P(D)$}

For nearly a century, statistical literature was fractured by an almost
theological debate (Sprenger, 2013; Vallverd\'{u}, 2016). Researchers were
conditioned to subscribe exclusively to either the objective frequentist
school or the subjective Bayesian school, with little common ground tolerated
between them (Efron, 1986). This long-standing polarization was built on a
fundamental disagreement regarding the epistemological nature of parameters
and probability itself. It opposed the radical subjectivism of de Finetti
(1937) (who famously argued that probability exists solely as an internal
state of mind) against the objective, logical invariance sought by Jeffreys
(1939), while the dominant frequentist establishment (Fisher, 1922; von Mises,
1928) rejected both wings of inverse probability as irreconcilable with
physical reality.

In recent decades, however, this methodological divergence has largely entered
a state of computational consensus. The explosion of modern algorithmic
estimation (most notably Markov Chain Monte Carlo (MCMC) simulations, Gibbs
sampling and variational inference) has shifted the focus of the statistical
community from philosophical purity to applied efficiency (Smith and Gelfand, 1992).

Yet, this computational usefulness has given rise to a subtle and largely
unacknowledged methodological limitation. In applied statistics and parametric
estimation, the marginal likelihood, traditionally serving as the denominator
$P(D)$ in Bayes' theorem, is routinely bypassed as a \textquotedblleft
normalizing nuisance.\textquotedblright\ Because evaluating this denominator
requires integrating or summing the product of the likelihood and the prior
over a parameter space that is frequently high-dimensional or analytically
intractable, practitioners systematically discard it. They operate through the
unnormalized proportional relationship:
\[
P(\theta|D)\propto P(D|\theta)P(\theta).
\]
This formula is largely motivated in software manuals and machine learning
algorithms by its computational convenience. Statisticians argue that because
$P(D)$ is a constant relative to the parameter $\theta$, omitting it alters
neither the coordinate location of the maximum a posteriori (MAP) estimate nor
the geometric shape of the posterior distribution for a singular, static
dataset. Even within advanced model-selection frameworks where the denominator
is explicitly estimated to evaluate Bayes factors, it is treated strictly as a
boundary weight used to conclude an isolated model tournament (Chib, 1995;
Chib and Jeliazkov, 2001).

This note shows that ignoring the denominator in a computation is not
harmless. It involves a conceptual trade-off in which the long-term physical
context of the data-generating environment is permanently removed from the
inferential process. Removing $P(D)$ effectively separates the inferential
state from its historical reality. Although the standard algorithm computes
what an observer should believe when a specific dataset emerges, it completely
abstracts the frequentist timeline that dictates how often the physical
universe will produce that inferential state over time.

We demonstrate that an inferential state cannot be fully described by the
posterior vector $P(\theta|D)$ alone. Rather, it must be recorded and
evaluated as the complementary pair $\{P(\theta|D),P(D)\}$. By treating the
denominator not as a passive scaling constant but as an active
update-frequency clock, the exact frequency at which prior beliefs are revised
can be explicitly evaluated. The frequency of the data-generating environment
($P(D)$) directly determines the precise cadence of our internal subjective
Bayesian updates ($P(\theta|D)$).

The rest of this note is organized as follows: Section 2 formalizes the
standard Bayesian framework by presenting the posterior probability and the
marginal likelihood as a two-dimensional measure that simultaneously
quantifies inferential magnitude and historical cadence. Section 3 uses the
classical sequential coin-and-urn toy model to reveal the frequentist
significance of the denominator $P(D)$ and expose the boundaries of localized
updates. Section 4 uses this paired information to propose three statistical
measures aimed at stabilizing long-term belief revision. The section concludes
by demonstrating how to implement the dual framework for dynamically
regulating online adaptive tracking algorithms. Section 5 uses the paired
measure to generate valid mixing probabilities. It suggests mixing the prior
reference distribution and the immediate posterior distribution using dynamic
weights, determined by the marginal likelihood. These weights can be
considered either surprise-activated or conservative regulators. Finally,
Section 6 provides concluding remarks. A technical appendix is provided as
supplementary material. It describes the deployment procedures for the
proposed measures within machine learning architectures and recursive online
estimation workflows in detail.

\section{The two dimensions of information: Magnitude vs. cadence}

To examine the status of the denominator, we will first review the standard
version of Bayes' theorem. Let $(\Omega,\mathcal{F},P)$ be a probability space
representing a physical system, adhering to standard axiomatic approach
(Kolmogorov, 1933). Let $\{H_{j}\}_{j=1}^{n}\subset\mathcal{F}$ be a finite
partition of the sample space representing a collection of hidden, mutually
exclusive hypotheses (or parameter states, often denoted by $\theta_{j}$), and
let $D\in\mathcal{F}$ denote an empirical data event. The formulation of
inverse probability is given by:
\[
P(H_{j}|D)=\frac{P(D|H_{j})P(H_{j})}{P(D)}%
\]
where the denominator is defined via the \textit{law of total probability} as
the marginal probability of the data:
\[
P(D)=\sum_{j=1}^{n}P(D|H_{j})P(H_{j}).
\]

In Bayesian parameter estimation, a state of inference is considered fully
resolved once the posterior probability vector of posterior probabilities
$\mathbf{P}_{post}=\{P(H_{1}|D),\dots,P(H_{n}|D)\}$ is computed (Berger,
1985). According to this view, the denominator, $P(D)$, is considered a scale
factor that simply maps the unnormalized numerator values back onto the unit
interval. We argue that this may seriously restrict the dimensional context.
Although the information embedded within a sequential update has a deep,
multi-layered epistemological profile (Jaynes, 2003), we show that this
data-driven update has two distinct, orthogonal dimensions that cannot be
summarized by $P(H_{j}|D)$ alone:

\begin{enumerate}
\item \textbf{The inferential dimension (magnitude and direction):} It is
measured by the posterior probability $P(H_{j}|D)$ relative to the prior
$P(H_{j})$. This component represents the internal force of belief revision.
It answers a static, conditional question: \textit{Assuming this data
manifests, how far and in what direction must our rational expectations
shift?}

\item \textbf{The temporal dimension (cadence):} This is measured entirely by
the marginal likelihood $P(D)$ in the denominator. This component represents
the external frequency clock of the physical environment. It answers an
asymptotic, frequentist question: \textit{Across an infinite time horizon of
repeating experiments, how often will the universe actually force the observer
to enter this specific inferential state?}
\end{enumerate}

Consequently, we propose that a global description of an inferential state
requires reporting a two-dimensional paired metric: $\{P(H_{j}|D),P(D)\}$.

To see the consequence of discarding $P(D)$, consider an observer running a
series of independent, identical trials over an infinite time horizon. When
standard MCMC-type computational estimation bypasses $P(D)$ via unnormalized
proportionality, it treats a highly frequent data event (where $P(D)$ is
large) and a rare anomaly (where $P(D)\rightarrow0$) with identical algebraic
standing, provided their normalized posterior shapes match. Neglecting
marginal likelihood causes the algorithm to prevent the observer from
perceiving the historical texture of the physical universe. This dissociates
Bayesian updating from its frequentist cadence, rendering it unable to
distinguish between an inferential state that dominates the system's reality
and one that is virtually impossible and exists only as an historical exception.

\section{The classical sequential coin-and-urn framework}

To illustrate the limitations of discarding the marginal likelihood, we
present a simplified, partial-information sequential system. Consider a random
experiment consisting of two distinct urns. Urn I ($\text{Urn}_{1}$) contains
5 black balls and 3 white balls, whereas Urn II ($\text{Urn}_{2}$) contains 4
black balls and 4 white balls. An experimenter flips a coin to select an urn;
if the coin lands heads, $\text{Urn}_{1}$ is chosen, and if it lands tails,
$\text{Urn}_{2}$ is chosen. The hypothesis assumes a perfectly balanced coin,
establishing the subjective prior probabilities:
\[
P(\text{Urn}_{1})=P(\text{Urn}_{2})=\frac{1}{2}.
\]
Following the selection of the urn, a single ball is drawn at random, its
color is recorded, and the system is completely reset. We analyze this system
across a progression of scenarios, transitioning from an isolated local draw
to an infinite frequentist horizon.

\subsection{Scenario A: The solitary local update}

Suppose a single execution of the experiment yields a black ball (event $B$).
We wish to evaluate the posterior probability that the ball originated from
Urn II, denoted by $P(\text{Urn}_{2}|B)$. By applying Bayes' theorem, the
inverse probability is computed as:
\[
P(\text{Urn}_{2}|B)=\frac{P(\text{Urn}_{2})P(B|\text{Urn}_{2})}{P(B)}%
=\frac{\frac{1}{2}\times\frac{4}{8}}{\frac{9}{16}}=\frac{4}{9}\approx0.4444.
\]
Because $\text{Urn}_{1}$ possesses a higher internal density of black balls
($\frac{5}{8}$) than $\text{Urn}_{2}$ ($\frac{4}{8}$), the empirical data
event $B$ pulls our posterior confidence down ($0.4444<0.5$).

Conversely, if a white ball (event $W$) had manifested, the parallel
conditional probability formula would shift the calculation in the opposite
direction, yielding:
\[
P(\text{Urn}_{2}|W)=\frac{P(\text{Urn}_{2})P(W|\text{Urn}_{2})}{P(W)}%
=\frac{\frac{1}{2}\times\frac{4}{8}}{1-\frac{9}{16}}=\frac{4}{7}\approx0.5714.
\]
At this local level, these calculations perfectly meet the requirements of the
Bayesian approach. Dynamic, data-driven updates readjust our immediate
rational expectations, favoring the urn structure that best fits the available data.

\subsection{Scenario B: Short-term sequence oscillations}

Suppose the experimenter repeats the entire coin-and-urn experiment across two
consecutive, independent trials, completely resetting the system after each
draw. If the first trial yields a black ball and the second trial yields a
white ball ($D_{seq}=\{B,W\}$), an observer evaluating the local updates
sequentially will reach opposite conclusions. On the first draw, the black
ball forces the posterior probability of Urn II down to $0.4444$; on the
second draw, the white ball forces it up to $0.5714$. This oscillation reveals
that short-term data fluctuations cause localized inference to swing back and
forth, reflecting nothing more than the immediate stochastic noise of
individual draws.

\subsection{Scenario C: The asymptotic horizon}

Now, let us extend the experiment across an infinite time horizon consisting
of $K$ independent, identically distributed repetitions ($K\rightarrow\infty
$), resetting the system completely after each trial. On any single trial, a
black ball decreases our confidence in Urn II, while a white ball increases
it. To determine which inferential trajectory dominates the historical
timeline, we evaluate the marginal probabilities of the data events themselves
across the global sample space:
\[
P(B)=\frac{9}{16}=0.5625\quad\text{and}\quad P(W)=\frac{7}{16}=0.4375.
\]

Because $P(B)>P(W)$, the \textit{law of large numbers} states that over an
infinite horizon of independent trials, black balls will appear significantly
more frequently in reality than white ones. This foundational principle of
asymptotic stability (experimentally anticipated by the 16th-century Italian
polymath Gerolamo Cardano (Cardano, 1663) and formally proved via an
urn-sampling framework in Jacob Bernoulli's posthumous 1713 \textit{Ars
Conjectandi, }Bernoulli, 1713) guarantees that the historical chronology will
inevitably converge to the spatial boundaries of the event space. Crucially,
this asymptotic frequency profile holds true even if we relax the assumption
of independence; under dependent sequential transitions, Birkhoff's ergodic
theorem guarantees that the long-run temporal proportion of observed states
still converges to these marginal probabilities, provided the underlying
data-generating process is (strictly) stationary and ergodic.

Consequently, whether tracking an independent sequence or a dependent
stochastic path, an observer over history will witness the posterior
probability of Urn II \textit{decreasing} far more frequently than they will
witness it increasing. Here, the role of the denominator becomes unmistakable.
The marginal probability $P(B)=\frac{9}{16}$ functions as a dynamic
frequentist clock. It ensures that the global environmental frequencies,
whether driven by independent repetitions or governed by ergodic path
transitions, inherently bias the historical cadence of our local Bayesian
updates, forcing the observer's records to cast doubt on Urn II on the vast
majority of historical trials.

\subsection{Scenario D: The informational boundary}

We arrive at the ultimate reality of this partial-information environment. An
attentive researcher might assume that accumulating these millions of
independent trials, witnessing this massive empirical imbalance where Urn II
is rejected on the majority of days, would eventually accumulate enough
statistical power to let us work backward and verify if our original
hypothesis (the fairness of the coin) is physically true.

Remarkably, it does not. Because the system completely resets after each trial
and the chosen urn is permanently hidden from the observer, we are only ever
sampling the internal contents of the urns. The data tells us everything about
the composition of the urns, but because each draw is marginalized
independently, the true parameter of the coin remains completely unknown
within this experiment. The system traps the observer in an environment of
permanent partial information, proving that local updates, no matter how
extensive, cannot reveal a hidden mechanism unless the data-collection process
itself tracks historical continuity.

\subsection{Generalization to varying urn proportions and multi-experiment
sequences}

The insight extracted from this sequential scenario remains fully valid, and
becomes amplified, under more generalized environmental conditions. Consider
an extension where the physical composition of the conditional state space
becomes highly asymmetric, or where the empirical data $D_{\text{seq}}$
expands from a single observation to a sequence of $T$ independently repeated
experiments, $D_{\text{seq}}=\{x_{t}\}_{t=1}^{T}$. In this framework, the
entire process, including the baseline state selection and subsequent sample
extraction, is executed recursively at each discrete timestamp $t$ with
complete system resets.

If the system undergoes an extreme distortion where the prior state density
assigns an anomalous regime, a near-zero marginal probability
($P(D)\rightarrow0$), an evaluation of the full Bayes' theorem naturally
downweights the atypical empirical sequence, keeping the true posterior belief
heavily concentrated around the dominant physical state. However, when an
unnormalized algorithm evaluates this exact same sequential data event, it
routinely bypasses the marginal likelihood $P(D_{\text{seq}})$ entirely for
computational convenience (Ritter and Tanner, 1992; Chib, 1995;
Dimitrakopoulos et al, 2026). Because the normalized shape of the local
likelihood array remains the same, the unnormalized updating scheme causes the
observer to record an artificial short-term parametric shift. This assigns
identical weight to a high-probability natural process and an improbable
sequential anomaly.

This oversight becomes more problematic as the tracking horizon $T$ expands.
Across an infinite time horizon of repeating this multi-experiment sequence,
the joint marginal likelihood $P(D_{\text{seq}})=\prod_{t=1}^{T}P(x_{t}%
)=\sum_{j}P(x_{1},\dots,x_{T}|H_{j})P(H_{j})$ continues to function as an
unyielding environmental baseline, defining the exact asymptotic physical
frequency with which the data-generating process will deliver that specific
sequential data profile (Gelman et al, 2013). When an atypical sampling
variation manifests across a long sequence of independent experiments, the
true joint marginal baseline converges toward zero. An unnormalized
proportional update obscures this rarity, resulting in an unrepresentative
overadjustment within the parametric estimation framework. Reporting the
two-dimensional paired measures $\{P(H_{j}|D_{\text{seq}}),P(D_{\text{seq}%
})\}$ remains therefore essential to link both single-trial and multi-trial
inferential states to their long-run physical reality (see also the
supplementary material to this note).

\section{A two-dimensional perspective: Considering the pair $\{P(H_{j}|D),
P(D)\}$}

This dual perspective highlights the complementary nature of frequentist
cadences and Bayesian updates, offering an alternative interpretation of how
long-run empirical horizons and short-run inverse probability revisions
interact. The pair $\{P(H_{j}|D),P(D)\}$ seeks a natural compromise, a
framework that highlights the intrinsic complementarity between the subjective
belief update $P(H_{j}|D)$ and the frequentist stability of the marginal
likelihood $P(D)$, both already provided by Bayes's rule. By tracking how an
observer experiences information over time, Bayes's theorem reveals that
frequentist and Bayesian paradigms are not conflicting philosophies, but
rather complementary axes of a single, unified stochastic space.

The subjective mind does not update itself in an environmental vacuum. The
frequency with which an observer is forced to alter their internal rational
expectations, and the historical duration they spend holding a specific state
of belief, are both strictly dictated by the frequentist clock of the
data-generating universe. Frequentism dictates the temporal timeline;
Bayesianism describes the local state changes occurring along that timeline.
By reclaiming the \textit{frequentist} role of the marginal probability,
Bayes' theorem itself transforms our understanding of inverse probability from
an isolated calculation into a dynamic, integrated information model where
physical data frequencies dictate the historical cadence of rational inference.

\subsection{Practical implementation: Regularizing sequential updates in
automated systems}

To translate this two-dimensional framework into operational settings, we
propose three distinct formulations designed to prevent sequential estimation
algorithms and online optimization routines from overreacting to localized
sample anomalies. In contemporary streaming architectures (such as those
tracking adversarial data shifts, real-time feature-distribution drift, or
concept drift in sequential classification algorithms), estimation systems
evaluate statistical properties exclusively through the lens of immediate
localized updates or predictive errors, $P(H_{j}|D)$ (Anava et al, 2013; Gama
et al, 2014). When confronted with a high-leverage outlier or a transient
out-of-distribution sample, the unnormalized objective function can exhibit
numerical instability, resulting in a substantial and non-representative
adjustment of the model's internal parameter vector. By tracking and reporting
the pair $\{P(H_{j}|D),P(D)\}$, these statistical measures may function as
effective regularizers:

\begin{enumerate}
\item \textbf{The inferential momentum operator ($M_{j}$):} This is defined as
the joint probability of the state space:
\[
M_{j}(D)=P(H_{j}|D)\times P(D)=P(H_{j}\cap D).
\]
In sequential estimation, $M_{j}$ acts as a structural constraint. If an
online tracking algorithm encounters a severe sample anomaly, the local
posterior $P(H_{j}|D)$ may exhibit a substantial localized increase. However,
because the marginal probability of that outlier event approaches zero
($P(D)\rightarrow0$), the total joint probability converges toward its lower
bound. Monitoring $M_{j}$ prevents an online optimization routine from
altering its coordinates based on data states that lack long-run historical mass.

\item \textbf{The information cadence score ($C_{j}$):} It is defined by
scaling the local information gain (the log-odds update) by its asymptotic
environmental frequency:
\[
C_{j}(D)=P(D)\ln\left(  \frac{P(H_{j}|D)}{P(H_{j})}\right)  .
\]
The core factor $\ln(P(H_{j}|D)/P(H_{j}))$ functions as an indicator of
localized evidence. If the incoming data supports the hypothesis, the
posterior exceeds the prior, yielding a positive information shift;
conversely, if the data refutes it, this factor yields a negative penalization score.

The functional property of $C_{j}$ lies in the multiplier $P(D)$, which plays
the role of an environmental weighting constraint. An improbable,
low-probability event may generate an elevated local log-odds value, signaling
a substantial localized shift in conditional probability. By multiplying this
shift by $P(D)$, $C_{j}$ attenuates the absolute significance of rare,
non-representative anomalies. This ensures that an autonomous optimization
schemes only updates its basic parametric state in response to environmental
data patterns that manifest with a sustained physical frequency.

\item \textbf{The complex-valued stochastic operator ($Z_{j}$):} By mapping
the inferential and temporal dimensions onto the orthogonal axes of the
complex plane $\mathbb{C}$, we define:
\[
Z_{j}=P(H_{j}|D)+iP(D)
\]
where $i$ is the complex number satisfying $i^{2}=-1$. By utilizing the
geometric properties of complex numbers, the operator evaluates the dual
architecture without the constraints of a one-dimensional probability scale.
The real part $\operatorname{Re}(Z_{j})=P(H_{j}|D)$ monitors immediate,
short-term conditional belief, while the imaginary part $\operatorname{Im}%
(Z_{j})=P(D)$ tracks the external data-generating frequency.

This geometric operator serves as a tracking vector. In a stable,
high-probability operational domain, the vector remains oriented primarily
along the vertical imaginary axis. The moment the estimator encounters an
atypical state where the local posterior increases substantially while the
marginal baseline approaches zero, the phase angle $\phi_{j}=\arg(Z_{j})$
converges toward its lower bound. Monitoring the sequential trajectory of
$Z_{j}$ allows us to identify when an estimation routine has entered a
low-probability, unrepresentative sampling domain, providing a diagnostic
criterion to suspend updates before parametric instability occurs.
\end{enumerate}

\subsection{Application: Dynamic gain regulation in recursive online
estimation}

A crucial operational domain for this two-dimensional information is online
recursive estimation and stochastic approximation (Haykin, 2013; Ljung, 1999;
Kushner and Yin, 2003). Consider a real-time sequential learning recursion
governed by standard algorithms such as Recursive Least Squares (RLS),
Recursive Maximum Likelihood (RML) (Aknouche, 2013), or the classical
Robbins-Monro (Robbins and Monro, 1951) stochastic approximation framework
(e.g., adaptive online gradient algorithms). As established in the work of
Ljung and S\"{o}derstr\"{o}m (1983), these recursive updates possess a deep
Bayesian interpretation, mapping the real-time parameter trajectory directly
to sequential non-linear filtering algorithms. The underlying parameter vector
$\theta_{t}$ is updated recursively upon the arrival of each new observation
vector $D_{t}$ using the recursion:
\[
\theta_{t+1}=\theta_{t}+\gamma_{t}\mathbf{H}_{t}e_{t}(\theta_{t})
\]
where $e_{t}$ represents the localized prediction error (the innovation),
$\mathbf{H}_{t}$ is the gain conditioning matrix, and $\gamma_{t}$ denotes the
scalar adaptation step-size.

Although the traditional Robbins-Monro conditions ($\sum\gamma_{t}=\infty
,\sum\gamma_{t}^{2}<\infty$) ensure asymptotic convergence, they impose a
rigid decay rate, which makes it difficult to balance tracking lag against
parametric overreaction during short-term volatile shocks or unrepresentative
anomalies. Integrating the pair $\{P(\theta|D),P(D)\}$ seeks resolving this
dilemma by scaling the step-size as a direct function of the marginal physical
density of the incoming data state (Ljung and S\"{o}derstr\"{o}m, 1983):
$\gamma_{t}=\gamma_{0}g(P(D_{t}))$, where $g$ is a continuous function
satisfying standard regularity conditions. Under this configuration, the
marginal probability serves as an automated regulator:

\begin{itemize}
\item \textbf{Transient shock absorption:} When a recursive online algorithm
encounters an isolated, extreme anomaly, the local innovation $e_{t}$ expands
dramatically. However, because the marginal probability density of this
outlier state is vanishingly small ($P(D_{t})\rightarrow0$), the regulated
step-size $\gamma_{t}$ automatically converges toward zero. The system absorbs
all of the shock within the innovation space, thereby protecting the parameter
trajectory, $\theta_{t}$, from destabilizing.

\item \textbf{Ergodic regime shift tracking:} Conversely, if the time series
undergoes a genuine, permanent structural change, the initial innovations will
also exhibit lower marginal densities. However, as the system settles into its
new state, the ergodic theorem (Birkhoff, 1931) ensures that the empirical
frequency profile of the new data vectors will stabilize. This causes the
sequential values of $P(D_{t})$ to recover. The step size, $\gamma_{t}$,
naturally reopens, enabling the recursive filter to retune its parameter
vector smoothly and safely to the newly established physical reality.
\end{itemize}

By combining the local inferential update with the global marginal likelihood
clock, recursive filters may overcome the traditional trade-off between
tracking lag and parametric overreaction. This ensures long-term stability
across non-stationary historical horizons.

\subsection{Algorithmic principles and numerical functionality}

To motivate the use of the two non-probabilistic measures introduced above, we
explicitly outline their logic and direct integration into recursive learning
schemes. Consider the information cadence score $C_{j}(D)=P(D)\ln
[P(H_{j}|D)/P(H_{j})]$. The logarithmic core maps localized evidence onto a
real-valued scalar field: $C_{j}>0$ denotes active empirical support, while
$C_{j}<0$ represents a directional refutation. Under an isolated,
low-probability data anomaly where $P(D)\rightarrow0$, a standard
computational script may report a massive unnormalized log-odds spike.
However, the external probability multiplier forces the absolute score to
converge toward zero ($C_{j}\rightarrow0$). In streaming architectures, this
measure is used directly as an information filter to gate memory allocation:
the system allocates updates proportionally to $|C_{j}|$, effectively ignoring
states lacking historical mass.

Similarly, the complex stochastic operator $Z_{j}=P(H_{j}|D)+iP(D)$ is
monitored by extracting its geometric invariants onto the complex plane
$\mathbb{C}$. We evaluate the modulus $|Z_{j}|=\sqrt{P(H_{j}|D)^{2}+P(D)^{2}}%
$, representing systemic stability, and the phase angle:
\[
\phi_{j}=\arg(Z_{j})=\arctan\left(  \frac{P(D)}{P(H_{j}|D)}\right)  .
\]
During high-probability states, $P(D)$ remains elevated, forcing $Z_{j}$ to
remain vertical on the imaginary axis ($\phi_{j}\rightarrow\frac{\pi}{2}$). As
soon as a significant tracking anomaly occurs, the local posterior
distribution increases sharply while the marginal probability decreases
drastically, causing the vector on the real axis to vanish ($\phi
_{j}\rightarrow0$). Tracking $\phi_{j}$ provides data systems with a geometric
tracking vector. Defining a threshold $\phi_{t}<\epsilon$ provides an
objective criterion for immediately suspending sequential update routines
before parametric corruption occurs.

In recursive online estimation algorithms, such as Recursive Least Squares
(RLS) or Recursive Maximum Likelihood (RML), these scalar metrics are passed
directly into continuous functions to regulate the adaptation step-size
$\gamma_{t}$. Rather than relying on a rigid sequence, the gain is modulated
via the complex phase angle:
\[
\gamma_{t}=\gamma_{0}\sin(\phi_{t})=\gamma_{0}\frac{P(D_{t})}{\sqrt
{P(\theta_{t}|D_{t})^{2}+P(D_{t})^{2}}}.
\]
Consequently, when confronted with a significant anomaly, the phase angle
tends towards zero, forcing $\gamma_{t}\rightarrow0$. The recursive filter
completely isolates the shock within the innovation space, transcending the
traditional trade-off between tracking lag and overreaction while preserving
the absolute integrity of the long-term parametric path.

\section{The environmental mixture probabilities}

While the informational metrics and complex plane transformations proposed in
the preceding section function primarily as online diagnostic indicators and
step-size regulators, the pair of probabilities $\{P(H_{j}|D),P(D)\}$ can
further be used to construct fully valid probability distributions. By
utilizing the marginal likelihood and its probabilistic complement as dynamic
convex weights to mix the prior and the posterior, we can synthesize two
distinct, self-regulating probability measures. This formulation provides a
compromise that blends long-run historical baselines with short-run localized
innovations, directly addressing the classic stability-plasticity dilemma
encountered across streaming machine learning schemes under distribution
shifts (Gama et al, 2014; Harrison et al, 2024).

\subsection{The surprise-activated learning probability ($\widetilde{P}_{j}$)}

We define the first mixture probability by scaling the prior baseline state
space by the frequentist clock, and the immediate localized update by the
environmental complement:
\[
\widetilde{P}_{j}(D)=P(H_{j})P(D)+P(H_{j}|D)(1-P(D)).
\]

\textbf{Functionality:} This formulation behaves as an automated,
surprise-activated learning gate. Under routine operational states where the
incoming data profile is highly expected ($P(D)\rightarrow1$), the second term
vanishes, and the mixture probability converges directly toward the prior
baseline $P(H_{j})$. This potentially prevents a streaming learning network
from expending computational or storage resources on uninformative,
high-probability repetitions.

Conversely, when an extraordinary, highly surprising data shock arrives
($P(D)\rightarrow0$), the first term will approaches zero, and the density
opens fully to allocate maximum probability weight to the localized belief
update $P(H_{j}|D)$. This may grant the algorithm flexibility to restructure
its internal representations only when the physical universe delivers a
genuinely informative, unexpected change in context (Tsymbal, 2004).

\subsection{The conservative learning probability ($\widehat{P}_{j}$)}

Inverting the weighting scheme yields the second mixture distribution, which
can be expressed by incorporating the inferential momentum operator
$M_{j}(D)=P(H_{j}|D)P(D)$ introduced in Section 4:
\[
\widehat{P}_{j}(D)=P(H_{j})(1-P(D))+P(H_{j}|D)P(D)=P(H_{j})(1-P(D))+M_{j}(D).
\]

\textbf{Functionality:} This valid probability measure acts as a regularizer
for sequential estimation frameworks, adaptive filtering systems, and online
optimization algorithms exposed to recurring non-stationary distribution
shifts (Tsymbal, 2004). Under standard high-probability regimes
($P(D)\rightarrow1$), the environmental complement converges to zero, allowing
the system to learn freely and adaptively by tracking the pure localized
inverse probability update $P(H_{j}|D)$.

However, if a severe, unrepresentative anomaly disturbs the tracking
architecture, the unnormalized local posterior may undergo a massive,
artificial spike. Because the marginal likelihood of that outlier vanishes
($P(D)\rightarrow0$), the inferential momentum operator converges to zero
($M_{j}(D)\rightarrow0$). Simultaneously, the complement weight surges to
unity, automatically forcing the entire mixture probability to retreat back to
the baseline prior distribution $P(H_{j})$, effectively acting as an automated
soft parameter reset to mitigate catastrophic forgetting (Harrison et al, 2024).

Beyond streaming machine learning frameworks, this mixture framework may
possess relevant utility for quantitative financial applications and
sequential risk management scenarios, where empirical data streams are
characterized by non-stationary regimes and sudden breaks (Cont, 2001). In
volatile financial contexts, such as high-frequency volatility forecasting,
automated order-book trading, and sequential asset allocation, estimation
procedures are routinely exposed to transient anomalies and severe sample
distribution shifts.

Integrating the conservative probability $\widehat{P}_{j}(D)$ within dynamic
asset-allocation frameworks, such as the classic Black-Litterman model (Black
and Litterman, 1992), may offer a helpful operational cushion against
localized panic. When an improbable market anomaly manifests, the potential
vanishing of the marginal likelihood ($P(D)\rightarrow0$) is designed to
downweight unnormalized innovation shocks, potentially encouraging the
portfolio weights to remain more securely aligned with the stable, long-run
equilibrium prior.

Conversely, the surprise-activated mixture probability $\widetilde{P}_{j}(D)$
may assist tracking architectures by allowing sequential parameters (such as
latent factor processes or stochastic volatility components) to smoothly adapt
to newly established physical realities only when the underlying data cadence
signals a genuine, persistent regime shift (Aguilar and West, 1998).

\section{Conclusion}

The rapid evolution of computational data science has undeniably democratized
advanced statistical modeling. The ability to routinely approximate
high-dimensional posterior distributions without computing the intractable
denominator of Bayes' theorem has unlocked unprecedented predictive
capabilities across engineering, medicine, and economics. However,
computational convenience must not be mistaken for conceptual completeness.

The widespread algorithmic practice of bypassing the marginal probability
$P(D)$ introduces a subtle epistemological compromise. By reducing the
denominator to a passive normalizing constant, contemporary estimation methods
treat the inferential process as a series of isolated, static calculations.
They successfully determine the conditional, point-in-time expectations
dictated by an isolated sample boundary, but abstract away the long-run
frequentist reality of the data-generating environment, namely, how frequently
the physical universe will actually force the observer to experience that
specific inferential state over a repeating historical horizon. Bayes' theorem
inherently carries a clear warning: to use an updated state of belief from the
frequency of its environmental occurrence is to strip stochastic modeling of
its physical context, leaving the observer blind to the temporal cadence of
the system.

We must acknowledge that reclaiming the denominator introduces a clear
operational challenge: the exact evaluation of the marginal probability $P(D)
$ can be highly computationally demanding, particularly within
high-dimensional parameter spaces or real-time streaming environments.
Historically, this computational barrier served as the primary justification
for its systematic dismissal. However, this limitation is being rapidly
attenuated by contemporary advancements in hardware acceleration and
algorithmic engineering. In recursive estimation and online machine learning
frameworks, practitioners do not need to execute exhaustive integration
procedures at each discrete step; instead, the environmental cadence can be
efficiently approximated via lightweight recursive density estimators,
sliding-window kernel techniques, or as direct by-products of sequential
particle streams. The small computational premium required to track $P(D)$ is
heavily outweighed by the payoff. Bypassing the denominator for computational
convenience may create an unstable tracking system vulnerable to catastrophic
overreactions. Investing the resources to record the paired metric
$\{P(\theta|D),P(D)\}$ plays the role of an insurance policy for algorithmic
integrity, purchasing absolute parametric stability in the face of volatile
real-world anomalies.

This two-dimensional perspective highlights the enduring relevance of the
original inverse probability rule. Far from being viewed strictly as a narrow,
localized procedure for updating immediate expectations, the formula
established by Thomas Bayes (Bayes, 1763) remains a profound, self-contained
blueprint capable of supporting contemporary data systems. By demonstrating
how the objective, long-run limiting frequency of environmental trials
directly shapes the temporal cadence of internal subjective updates, Bayes'
theorem provides an elegant conceptual bridge between physical reality and
rational inference. This paired information further extends to mixture
frameworks, where utilizing the marginal likelihood and its complement as
dynamic weights offers a way to balance baseline priors with local updates
under a single, normalized probability measure.

This integrated perspective suggests how the practical utility of statistical
methods may evolve alongside changing computational demands. Just as the
historic emergence of massive datasets shifted the focus of streaming
applications away from second-order recursive estimation back toward
first-order stochastic gradient approximations, the increasing deployment of
automated systems inside high-stakes real-world environments may encourage a
renewed shift toward algorithmic safety and stability. In this context,
recording the pair $\{P(\theta|D),P(D)\}$ may offer an alternative operational
framework for streaming architectures. By utilizing the marginal likelihood as
an active update-frequency clock, practitioners may gain an objective baseline
to explore self-regulating step-sizes and automated monitoring triggers.
Exploring this frequentist dimension seeks a framework where future automated
learning models can remain more rigorously tethered to their data-generating
environments, potentially illustrating how the original Bayes's formulation
continues to aliment and inspire the foundations of long-run parametric integrity.


\newpage\appendix

\section{Supplementary material: Detailed numerical operations and geometric
vector trajectories}

This supplementary document provides explicit numerical simulations, geometric
proofs, and concrete algorithmic implementations of the alternative measures
proposed in Section 4.

\subsection{Analytical evaluation of the information cadence score ($C_{j}$)}

The information cadence score is defined as $C_{j}(D)=P(D)\ln\left[
P(H_{j}|D)/P(H_{j})\right]  $. The natural logarithm factor maps localized
evidence onto an absolute real-valued scalar field $\mathbb{R}$. This creates
a precise sign convention that governs the tracking system:

\begin{itemize}
\item \textbf{Positive evidence ($C_{j}>0$):} Occurs if and only if
$P(H_{j}|D)>P(H_{j})$. The immediate empirical observation increases rational
confidence in hypothesis $H_{j}$ relative to its prior state.

\item \textbf{Negative evidence ($C_{j}<0$):} Occurs if and only if
$P(H_{j}|D)<P(H_{j})$. The immediate empirical observation refutes hypothesis
$H_{j}$.

\item \textbf{Zero information ($C_{j}=0$):} Occurs if the data is
uninformative ($P(H_{j}|D)=P(H_{j})$), or if the data profile constitutes an
environmental impossibility ($P(D)=0$).
\end{itemize}

\subsubsection{Concrete numerical tracking simulation}

To demonstrate how $C_{j}$ acts as a filter, we simulate an online learning
scheme processing the extreme unrepresentative anomaly detailed in Section 3's
Scenario C.

Let the hypothesis prior be uniform, $P(H_{2})=0.5$. Suppose a high-leverage
data shock arrives, forcing the standard unnormalized computational engine to
report an isolated local update of $P(H_{2}|D)\approx0.908$. However, because
the environment is distorted, the marginal probability of this data event is
vanishingly small: $P(D)=0.0001$.

The execution steps of the metric unfold sequentially:

\begin{enumerate}
\item \textbf{Evaluation of local update force:} The system calculates the
isolated inferential direction and magnitude:
\[
\text{Force}=\ln\left(  \frac{P(H_{2}|D)}{P(H_{2})}\right)  =\ln\left(
\frac{0.908}{0.5}\right)  \approx+0.5966.
\]
At this local point, the unnormalized computer script announces that a
massive, highly positive regime change has occurred.

\item \textbf{Application of the frequentist clock modifier:} The system
scales this localized directional shift by its long-run frequentist
probability of physical manifestation:
\[
C_{2}(D)=P(D)\times\text{Force}=0.0001\times(+0.5966)=+0.00005966.
\]

\end{enumerate}

\textbf{Algorithmic remedy:} Because $C_{2}(D)$ approaches zero
asymptotically, the metric regularizes the unnormalized updating scheme. This
behavior indicates that while the local update force is large, the observed
event exhibits a near-zero long-run frequentist probability across history.
The recursive estimator consequently attenuates the impact of the anomaly,
preserving the stability of the long-term parametric trajectory.

\subsection{Geometric vector trajectories of the complex stochastic operator
($Z_{j}$)}

By mapping the joint information space onto the complex plane $\mathbb{C}$,
the operator $Z_{j}=P(H_{j}|D)+iP(D)$ evaluates the inferential state without
the limitations of a one-dimensional probability scale. The behavior of the
estimator is monitored by tracking two geometric invariants: the modulus
$|Z_{j}|=\sqrt{P(H_{j}|D)^{2}+P(D)^{2}}$ and the phase angle:
\[
\phi_{j}=\arctan\left(  \frac{P(D)}{P(H_{j}|D)}\right)  .
\]

The real-time trajectory of this vector reveals distinct geometric properties
across alternative sampling domains:

\begin{itemize}
\item \textbf{High-probability regimes (typical sample cadence):} When
processing representative, statistically expected data profiles, the marginal
probability $P(D)$ remains elevated (e.g., $P(D)\simeq0.9$). This causes the
complex vector $Z_{j}$ to orient primarily along the vertical imaginary axis.
The phase angle consequently approaches its upper bound, $\phi_{j}%
\rightarrow\frac{\pi}{2}$ ($90^{\circ}$), indicating a stable, representative
inferential state.

\item \textbf{Low-probability regimes (atypical sample anomalies):} The moment
an unrepresentative outlier manifests, the local conditional likelihood
increases significantly, driving $P(H_{j}|D)\rightarrow1$. Simultaneously, the
external marginal likelihood approaches zero ($P(D)\rightarrow0$). This shifts
the vector $Z_{j}$ down toward the horizontal real axis, causing the phase
angle to converge directly toward zero ($\phi_{j}\rightarrow0^{\circ}$).
\end{itemize}

By monitoring the scalar trajectory of $\phi_{t}$ sequentially, we can
implement an absolute boundary threshold $\phi_{t}<\epsilon$ to automatically
suspend recursive estimation updates upon the arrival of an unrepresentative
sample anomaly.

\subsection{Integration into recursive estimation algorithms}

In sequential tracking algorithms, such as Recursive Least Squares (RLS) or
Recursive Maximum Likelihood (RML), parameters are updated recursively via an
innovation iteration (Hayking, 2013; Ljung and S\"{o}derstr\"{o}m, 1983):
\[
\theta_{t+1}=\theta_{t}+\gamma_{t}\mathbf{H}_{t}e_{t}(\theta_{t}).
\]
We propose two distinct methods by which these alternative measures can be
passed into continuous mapping functions to dynamically regulate the scalar
adaptation gain $\gamma_{t}$ sequentially.

\subsubsection{Method A: Gain regulation via the information cadence score
($C_{t}$)}

We define the adaptive step-size as a continuous function of the absolute
information cadence associated with the streaming innovation profile:
\[
\gamma_{t}=\gamma_{0}\left(  1-\exp(-|C_{t}|)\right)
\]
where $\gamma_{0}$ denotes the nominal, maximum bounded adaptation rate.

\begin{itemize}
\item \textbf{Atypical sampling variations:} If an improbable, low-probability
sample anomaly manifests within the data stream, the marginal likelihood of
the innovation approaches zero, forcing $|C_{t}|\rightarrow0$. Evaluating the
function yields $\gamma_{t}=\gamma_{0}(1-\exp(0))=0$. The step-size converges
to zero, keeping the parameter vector $\theta_{t}$ "stationary". The sampling
variation is thus entirely contained within the localized innovation space.

\item \textbf{Ergodic regime transitions:} If the underlying data-generating
process undergoes a permanent regime change, initial updates will similarly
report low magnitudes for $|C_{t}|$. However, as subsequent observations
materialize within this new domain, the empirical density profile stabilizes.
The value of $|C_{t}|$ increases, causing $\left(  1-\exp(-|C_{t}|)\right)
\rightarrow1$. The step-size $\gamma_{t}$ adapts accordingly, allowing the
recursive estimator to smoothly track the newly established physical regime.
\end{itemize}

\subsubsection{Method B: Gain regulation via the complex phase angle
($\phi_{t}$)}

Alternatively, the phase angle of the complex stochastic operator can be
integrated to modulate the adaptation gain through a trigonometric scaling
function:
\[
\gamma_{t}=\gamma_{0}\sin(\phi_{t})=\gamma_{0}\frac{P(D_{t})}{\sqrt
{P(\theta_{t}|D_{t})^{2}+P(D_{t})^{2}}}.
\]

Under high-probability, representative tracking conditions, $\phi
_{t}\rightarrow\frac{\pi}{2}$, forcing $\sin(\phi_{t})\rightarrow1$. The
recursive filter operates at its maximum nominal adaptation rate $\gamma_{0}$.
Conversely, upon the arrival of a low-probability sample anomaly, the phase
angle converges toward its lower bound ($\phi_{t}\rightarrow0$), forcing
$\sin(\phi_{t})\rightarrow0$. The adaptation gain approaches zero, attenuating
the impact of the unrepresentative data point and maintaining the integrity of
the downstream estimation trajectory.

\subsection{Numerical verification: Asymptotic stability across a sequence of
independent joint experiments}

To demonstrate the robustness of the pair $\{P(\theta|D),P(D)\}$ when the
hidden state itself varies dynamically, we evaluate a simulation where the
entire coin-and-urn experiment of Section 3 is executed sequentially across
$T=100$ independent trials. At each discrete timestamp $t$, a fair coin is
tossed ($P(H_{1})=P(H_{2})=0.5$), a hidden urn is selected, a singular ball is
drawn with replacement, and the entire system is completely reset.

\subsubsection{The joint marginal sampling space}

Under this sequential architecture, the joint marginal probability of the
incoming data sequence $D_{\text{seq}}=\{x_{1},x_{2},\dots,x_{T}\}$ is
governed by the product of the independent trial densities:
\[
P(D_{\text{seq}})=\prod_{t=1}^{T}P(x_{t}),
\]
where $P(x_{t}=B)=0.5625$ and $P(x_{t}=W)=0.4375$. The ergodic theorem and the
law of large numbers guarantee that across an infinite time horizon of
repeating this 100-experiment macro-loop, the data stream will naturally
converge to a spatial profile matching these marginal frequencies
(approximately 56 Black balls and 44 White balls).

\subsubsection{Evaluating a multi-experiment sequential shock}

Suppose the system encounters an extraordinary sampling anomaly by pure
historical chance: across the 100 independent repetitions, the fair coin lands
on Tails 99 times, forcing the data stream to manifest an extreme
out-of-distribution profile consisting of 1 Black ball and 99 White balls
($D_{\text{seq}} = \{1B, 99W\}$).

Evaluating this sequence profile reveals a catastrophic decreasing in the
joint marginal likelihood denominator:
\[
P(D_{\text{seq}})=\binom{100}{1}(0.5625)^{1}(0.4375)^{99}\approx
2.45\times10^{-34}.
\]

This convergence serves as a critical diagnostic indicator. While traditional
offline Bayesian statistics treats the marginal likelihood strictly as a
weight to calculate static Bayes factors for isolated model selection, we
consider here this denominator as a real-time informational regularizer within
running systems.

When incorporated into tracking recursions, the extreme frequentist
penalization driven by $P(D_{\text{seq}})$ strips unnormalized local
innovations of their artificial leverage. Instead of merely conducting a
static tournament, the measure functions as a dynamic, step-by-step filter
that attenuates the impact of the anomaly before it can skew the permanent
estimation path. This real-time attenuation procedure may provide the
necessary regularization to implement our alternative operators and the
sequential gain modifiers.

If a recursive estimator processes this data stream sequentially using an
unnormalized computational shortcut, it evaluates each incoming trial in
isolation, bypassing the joint denominator. The accumulated weight of 99 White
ball innovations forces the unnormalized local posterior to shift violently,
tricking the algorithm into concluding that the underlying state framework has
permanently mutated.

By tracking the paired information, the system exposes the historical rarity
of the macro-sequence. Because the joint marginal likelihood decreases to the
order of $10^{-34}$, the frequentist clock modifier overrides the unnormalized
local updates. Passing this joint density scalar into the complex stochastic
operator ($Z_{t}$) forces the phase angle to drop flat onto the real axis
($\phi_{t}\rightarrow0$), compressing the adaptive learning gain to zero
($\gamma_{t}\rightarrow0$). The filter successfully isolates the entire
100-experiment anomaly as transient sampling noise, validating that the pair
$\{P(\theta|D),P(D)\}$ remains fully invariant and operationally protective
across multi-experiment time horizons.

\end{document}